\documentclass[journal=jacsat,manuscript=article]{achemso}

\usepackage[version=3]{mhchem} 
\usepackage[utf8]{inputenc}
\usepackage{physics}
\usepackage{amsfonts}
\usepackage{amsmath}
\usepackage{bm}
\usepackage{amssymb}

\newcommand{\onlinecite}[1]{[\hspace{-1 ex} \nocite{#1}\citenum{#1}]} 

\author{Antoine Honet}
\affiliation{%
Department of Physics and Namur Institute of Structured Materials, University of Namur, Rue de Bruxelles 51, 5000 Namur, Belgium
}%
\email{antoine.honet@unamur.be}

\author{Luc Henrard}
\affiliation{%
Department of Physics and Namur Institute of Structured Materials, University of Namur, Rue de Bruxelles 51, 5000 Namur, Belgium
}%

\author{Vincent Meunier}%
\affiliation{%
 Department of Physics, Applied Physics, and
Astronomy, Rensselaer Polytechnic Institute, Troy, New York 12180, USA,
}%

\title{Exact and many-body perturbation solutions of the Hubbard model applied to linear chains.}

\keywords{Hubbard model, Green's functions, mean-field approximation, GW approximation, exact diagonalization, correlated systems, symmetry breaking}

\begin{document}


\begin{abstract}
  This study reports on the accuracy of the GW approximation for the treatment of the Hubbard model, compared to exact diagonalization (ED) results.  GW is part of the more general Green's functions approach used to develop many-body approximations (GFMBA). We show that, for small linear chains, the GW approximation corrects the mean-field (MF) approach by reducing the total energy as well as the magnetization obtained from the MF approximation. The GW energy gap is in a better agreement with ED, especially in even-number of atoms systems where, in contrast to the MF approximation, no plateau is observed below the predicted phase transition. In terms of density of states, the GW approximation induces quasi-particles and side satellites peaks via a splitting process of MF peaks. At the same time, GW slightly modifies the localization (\textit{e.g.}, edges or center) of the states. We also use the GW approximation results in the context of the L\"owdin's symmetry dilemma and show that GW predicts a paramagnetic-antiferromagnetic phase transition at a higher Hubbard parameter than MF does.
\begin{description}
\item[Keywords]
Hubbard model, Green's functions, mean-field approximation, GW approximation, exact diagonalization, correlated systems, symmetry breaking
\end{description}
\end{abstract}

\section{Introduction}

The Hubbard model has been introduced more than 50 years ago~\cite{hubbard_electron_1963} and is now a reference approach to describe correlations in various types of physical systems, ranging from electronic systems ~\cite{castro_neto_electronic_2009, yazyev_emergence_2010}, optical trapped and ultra-cold atoms~\cite{dutta_non-standard_2015, esslinger_fermi-hubbard_2010}, and quantum simulation~\cite{tarruell_quantum_2018}. It is thus a versatile description of the correlated systems that is the subject of many fundamental studies, ~\cite{tasaki_hubbard_1998,yepez_lecture_nodate, pavarini_physics_2017} even if it remains a simplified approach, containing only few parameters.

The Hubbard model is of particular importance to the description of magnetic properties of nano-systems~\cite{feldner_magnetism_2010,bullard_improved_2015, yazyev_emergence_2010}. Indeed, it includes the necessary interactions between electrons of opposite spins to predict the total spin as well as spin symmetry, which are key properties in the field of magnetic nano-materials and spintronics.

Despite its apparent simplicity, the Hubbard model is very difficult to solve. The exact treatment is performed by the so-called the exact diagonalization (ED)~\cite{lin_exact_1993, jafari_introduction_2008,kingsley_exact_2013, sharma_organization_2015} that requires switching the point of view from the one-body basis usually related to an independent electron picture to a many-body basis. While the ED technique is conceptually well understood, its applicability is limited because of its heavy numerical cost, due to the fact that the size of the Hilbert space grows rapidly with the number of particles. For example, its recent application to nano-graphene has been limited to $16$ carbon atoms only~\cite{feldner_magnetism_2010}.

In the electronic domain, a mean-field (MF) approximation is often introduced to approximate the Hubbard model~\cite{yazyev_emergence_2010,feldner_magnetism_2010,  claveau_mean-field_2014, bullard_improved_2015, raczkowski_hubbard_2020, zang_hartree-fock_2021}. This approximation allows to treat larger systems (up to thousands of atoms) because it can be resolved using a one-body basis set. However, the MF approach misses all the correlation effects introduced by the full Hubbard model. It is thus important to compare in details the MF results to ED or other advanced techniques (\textit{e.g.} quantum Monte Carlo or configuration interaction) ~\cite{feldner_magnetism_2010, raczkowski_hubbard_2020, joost_lowdins_nodate} to determine the validity of the MF approximation for different systems as a function of the model's parameter.

Alternatively, correlations can be re-introduced in the description of the electronic properties using a more tractable computational technique compared with ED, such as the Green's functions formalism, together with many-body approximations (GFMBA). The $GW$ approximation is one among a number of other GFMA approaches, such as the second-order or T-matrix approximations~\cite{stefanucci_nonequilibrium_2013, romaniello_beyond_2012} and it was first introduced by Hedin~\cite{hedin_effects_1970}. Recent reviews on $GW$ can be found in Refs.~\onlinecite{reining_gw_2018, golze_gw_2019}. The first one focuses mainly on the physical foundations of the $GW$ approximation while the second highlights results of current implementations in quantum chemistry. Additional reviews and textbooks also provide good insight about the $GW$ approximation~\cite{aryasetiawan__1998,
onida_electronic_2002,stefanucci_nonequilibrium_2013, martin_interacting_2016}.

Although known for a long time and widely used in quantum chemistry and \textit{ab initio} computations, it is not clear which part of the correlation many-body approximations include. It is not clear either to what extend, for which properties and for which systems, GW improves the MF performance compared with ED for a given range of parameter values. There has been some research devoted to this topic in the late 90's~\cite{schindlmayr_violation_1997,pollehn_assessment_1998,schindlmayr_spectra_nodate} but most of them focused on the density of states (DOS) and the total energy ($E_{tot}$), leaving unexplored the local density of states (LDOS) as well as the symmetry of the ground state. To address this shortcoming, the LDOS as well as density of the total states are investigated in this report.

A recent study~\cite{joost_lowdins_nodate} showed that the so-called L\"owdin's symmetry dilemma~\cite{lykos_discussion_1963} is not satisfactorily resolved using GFMBA, and more specifically in second-order Born approximation (SOA). L\"owdin's symmetry dilemma was first described within the MF approximation where, in some cases, solutions with unphysical spatial and spin symmetries are more stable. More precisely, the authors of Ref.~\onlinecite{joost_lowdins_nodate} highlight the case of one-dimensional (1D) single-orbital Hubbard chains with an even number of atoms for which the ED predicts a paramagnetic (PM) ground state for all values of the Hubbard parameters whereas MF predicts an anti-ferromagnetic (AFM) ground state above a critical Hubbard parameters value. If the PM symmetry is imposed in the MF algorithm, the total energy increases and deviates from the ED total energy. In other words, there is a phase transition predicted from PM to AFM in the MF approximation but not in the exact ED.  

In this paper, we investigate the difference in energy between PM and AFM states calculated in MF and in GW for a wide range of parameter value. The symmetry of the predicted states and their difference in energies are of importance for potential technological applications such as spintronics~\cite{bullard_improved_2015}. As a key result of this study, we show that the GW correction extends the parameter range for which the ground states display the correct spin symmetry compare to the MF approximation. We also show that the results from SOA for linear chains~\cite{joost_lowdins_nodate} remain valid within the GW approximation, leading to a L\"owdin's symmetry dilemma.

Analytical studies to aid in the interpretation of the reasons of success and failure of $GW$ and other approximations have been published in the literature~\cite{romaniello_self-energy_2009, romaniello_beyond_2012, di_sabatino_reduced_2015, tomczak_proprietes_2007, strunck_combining_nodate}. These analytical results are mainly obtained for the Hubbard dimer (a chain with 2 atoms). Here, we also provide new insight about their generalization to larger systems.

The rest of the paper is organized as follows: we first introduce the Hubbard model as well as the different approximations and formalisms (Green's functions, MF, GW and ED). We then study the improvement of GW approximation on global properties of the system, such as total energy, energy gap, density matrix or magnetization before switching to frequency-dependent quantities (including density of states and scanning tunneling spectroscopy simulations). The last section also discusses phase transition and how GW fares with respect to the L\"owdin's symmetry dilemma.

\section{Methods}

\subsection{Hubbard model}

 In this paper, we focus on a single-orbital Hubbard model for fermions that contains two parameters:
\begin{equation}
\hat{H}_{Hubbard } = - t \sum_{<ij>, \sigma}  (\hat{c}^\dagger_{i\sigma} \hat{c}_{j\sigma}  +c.c.) + U \sum_i \hat{n}_{i \uparrow} \hat{n}_{i \downarrow}
\label{Hubbard_ham}
\end{equation}
where $t$ is the hopping parameter, $U$ is the interaction (or Hubbard) parameter, $\hat{c}^\dagger_{i\sigma}$ (resp. $\hat{c}_{i\sigma}$) is a creation (resp. destruction) operator of an electron on atomic site $i$ with spin $\sigma$, and $\hat{n}_{i \sigma} = \hat{c}^\dagger_{i\sigma} \hat{c}_{i\sigma}$ is the density operator (of electron on atomic site $i$ and with spin $\sigma$). The $\langle$ $\rangle$ signs under the summation symbol indicate that the sum runs over all pairs of nearest-neighbor atomic sites.

The first term of the Hubbard Hamiltonian in eq.~\ref{Hubbard_ham} is the tight-binding Hamiltonian and can be written easily in the one-electron basis but the second term (Hubbard or interaction term) is the product of two density operators (consequently of four creation or destruction operators) such that it is called a two-body operator. It cannot be written in the one-electron basis, and a many-electron basis is required.

\subsection{Mean-field approximation}

The MF approximation consists of approximating the Hubbard Hamiltonian of eq.~(\ref{Hubbard_ham}) with one-body operators only, so that it can be written in the one-electron basis. In the single orbital-per-atom approach used here, the one-electron basis is of size $2N_{\rm at}$ since it consists of basis vectors of localised electrons on atomic sites with a given spin (up or down), responsible for the factor $2$ in the basis size. Doing so, the size of the basis is significantly reduced compared to the many-body case (see Exact diagonalization section).

In practice, the density operators ($\hat{n}_{i \sigma}$) are decomposed in the mean value of the operator ($n_{i \sigma}$) and the deviation ($\hat{n}_{i \sigma} - n_{i \sigma}$) from this mean value: $\hat{n}_{i \sigma} = n_{i \sigma} + (\hat{n}_{i \sigma} - n_{i \sigma})$. Then, products of density operators in eq.~(\ref{Hubbard_ham}) are expanded and products of deviations are dropped as they are expected to be small, leading to the mean-field Hamiltonian:
\begin{equation}
\begin{split}
   \hat{H}_{Hubb,MF} = & - t \sum_{<ij>, \sigma}  (\hat{c}^\dagger_{i\sigma} \hat{c}_{j\sigma}   +c.c.)+ U \sum_i ( n_{i \uparrow} \hat{n}_{i \downarrow} + n_{i \downarrow} \hat{n}_{i \uparrow}) \\
   &-  U \sum_i n_{i \uparrow} n_{i \downarrow}
\end{split}
\label{eq:ham_hubb_mf}
\end{equation}
and total energy:
\begin{equation}
E_{Tot, MF} = \sum_l E_l f_{\rm FD}(E_l)  
\label{eq:MF_Etot}
\end{equation}
where $E_l$ are the one-electron energies obtained by diagonalizing the Hamiltonian of eq.~(\ref{eq:ham_hubb_mf}), $f_{\rm FD}$ is the Fermi-Dirac and $n_{i \uparrow}$, and $n_{i \downarrow}$ are obtained from the eigen-vectors of the Hamiltonian of eq.~(\ref{eq:ham_hubb_mf}). More details about MF approximation can be found in Refs.~\onlinecite{bullard_improved_2015, yazyev_emergence_2010, stefanucci_nonequilibrium_2013}.

We emphasize here that, since the Hamitlonian~(\ref{eq:ham_hubb_mf}) depends on mean values of density operators, it has to be solved self-consistently starting from initial guesses for the mean values. Depending on the symmetry of the initial guess (\textit{e.g.}, PM or AFM), the self-consistent cycle can converge towards different (stable or metastable) states. The comparison of total energies associated with the states (eq.~(\ref{eq:MF_Etot})) is needed to identify the predicted ground state with the lowest total energy. Usually, the MF ground state is found by performing the MF self-consistent algorithm  several times with random initial states.

\subsection{GW approximation}

In order to account for a portion of the correlation, a Green's function based method to compute GW approximation on top of MF computations is used in this paper, following  Refs.~\onlinecite{joost_correlated_2019, honet_semi-empirical_2021}. It is based on the Feynman diagram expansion of the Green's functions explained, for example, in Ref.~\onlinecite{stefanucci_nonequilibrium_2013}. In this section, we first give a brief reminder of the Green's function formalism. We then present the main steps of the algorithm for a numerically approach and, finally, we give the main expressions to retrieve the useful quantities such as LDOS, total energy and STS simulated images.

In the Green's function formalism, we consider retarded (R), advanced (A), greater ($>$), and lesser ($<$) components of several quantities (Green's function ($G$), self energies ($\Sigma$) and screened potential ($W$)). We stress that these components are functions of real time instead of complex time for the total quantities. For definitions, properties and links between the components, we refer to chap.5 of Ref.~\cite{stefanucci_nonequilibrium_2013}. The different properties and links between components require to consider equations involving $\lessgtr$ components or equations involving $R/A$ components.

The retarded component of the Green's function ($G^{R} (\omega)$) obeys Dyson's equation:
\begin{equation}
    [G^{R} (\omega)]^{-1} = [G^{R}_0 (\omega)]^{-1} - \Sigma^{R} (\omega).
\label{eq:Dyson_eq}
\end{equation}
where $G^{R}_0 (\omega)$ is the non-interacting retarded Green's function and $\Sigma^{R} (\omega)$ is the retarded self-energy. All quantities in eq.~(\ref{eq:Dyson_eq}) are matrices in the one-electron basis (with dimension $2N \cross 2N$, $N$ being the number of sites).

The retarded and advanced non-interacting Green's functions in the GW approximation ($G^{R}_0 (\omega)$) are taken as the Green's functions of the MF solution, given by:
\begin{equation}
    G^{R/A}_{0, i\sigma, j\sigma'} (\omega) = \bra{i\sigma} (\omega - \hat{H}_{Hub,MF} \pm i\eta)^{-1} \ket{j\sigma'}
    \label{eq:non_int_G}
\end{equation}
where $\ket{j\sigma'}$ is the one-electron state containing one electron with spin $\sigma'$ on site $j$, $\bra{i\sigma}$ is the hermitian transpose of $\ket{i\sigma}$, and $\eta$ is a small real positive parameter.

The greater and lesser ($>$ and $<$ indices) components of the Green's functions are given by (the expressions remain valid for the exact and non-interacting Green's functions):
\begin{equation}
\begin{split}
    &G^<(\omega) = -f_{FD}(\omega-\mu) [G^R(\omega)-G^A(\omega)]\\
    &G^>(\omega) = \bar{f}_{FD}(\omega-\mu) [G^R(\omega)-G^A(\omega)]
\end{split}
\label{G_lesser_greater}
\end{equation}
with $\bar{f}_{FD}(\omega) = 1-f_{FD}(\omega)$ and $\mu$ the chemical potential.

Dyson's equation (eq.~(\ref{eq:Dyson_eq})) provides the exact Green's function ($G^{R} (\omega)$) from the non-interacting Green's function ($G^{R}_0 (\omega)$) and from the exact self-energy $\Sigma^{R} (\omega)$. But the exact self-energy is not known and approximations have to be made.

In the GW approximation, the self-energy is approximated as the product of the Green's function (G) and the screened potential (W):
\begin{equation}
    \Sigma^{\lessgtr} (t) = i W^{\lessgtr} (t)~\circ G^{\lessgtr} (t)
\label{Sigma_time}
\end{equation}
where $~\circ$ is the Hadamard product, or element-wise product between matrices.

$G^{\lessgtr} (t)$ in time domain is found by performing an inverse Fourier transform of $G^{\lessgtr} (\omega)$. The screened potential $W^R(\omega)$ obeys a Dyson-like equation:
\begin{equation}
    W^R(\omega) = W_0 + W_0 \chi^0(\omega) W^R(\omega)
\label{screened_W}
\end{equation}
where $W_0$ is the interaction matrix containing Hubbard $U$ term of Hamiltonian~(\ref{Hubbard_ham}) and $\chi^0(\omega)$ is the non-interacting susceptibility found by inverse Fourier transform of:
\begin{equation}
    \chi^0(t) = i \Theta(t) [G^>(t)\circ (G^<(t))^* - G^<(t)\circ (G^>(t))^*]
\label{chi0_time}
\end{equation}
where $\Theta(t)$ is the Heaviside step function.

With this \textit{proviso},  the greater and lesser components of the screened interaction of eq.~(\ref{Sigma_time}) are found from the relations between greater and lesser, and retarded screened interactions:
\begin{equation}
    \begin{split}
        &W^<(\omega) = 2i f_{\rm BE} (\omega - \mu) Im(W^R(\omega))\\
        &W^>(\omega) = 2i (f_{\rm BE}(\omega - \mu)+1)  Im(W^R(\omega))\\
    \end{split}
    \label{eq:W_less_gr}
\end{equation}
where $f_{\rm BE}$ is the Bose-Einstein distribution.

Finally, the retarded self-energy in frequency domain $\Sigma^{R} (\omega)$ needed in Dyson's equation~(\ref{eq:Dyson_eq}) is found by inverse Fourier transform of the time-domain retarded self-energy expressed in terms of greater and lesser self-energies:
\begin{equation}
\Sigma^R(t) = \Theta(t) [\Sigma^>(t) - \Sigma^<(t)].
\label{eq:sigma_R}
\end{equation}

The algorithm to find the GW Green's function self-consistently can thus be outlined as follows:
\begin{enumerate}
\item Initialize the Green's function to the non-interacting one (see eq.~(\ref{eq:non_int_G})): $G^{R}(\omega) = G^{R}_{0}(\omega)$
\item Fourier transform Green's function $G^{\lessgtr}(\omega)$ to get $G^{\lessgtr}(t)$
\item compute $\chi^0(t)$ using previous Green's function $G^{\lessgtr}(t)$ and eq.~(\ref{chi0_time})
\item Fourier transform the polarizability $\chi^0(t)$ to get $\chi^0(\omega)$
\item compute $W^R(\omega)$ and $W^{\lessgtr}(\omega)$ using $\chi^0_{j}(\omega)$ and equations~(\ref{screened_W}) and~(\ref{eq:W_less_gr})
\item Fourier transform the screened interaction $W^{\lessgtr}(\omega)$ to get $W^{\lessgtr}(t)$
\item compute $\Sigma^{\lessgtr} (t)$ and $\Sigma^R(t)$ using $W^{\lessgtr}_j(t)$, $G^{\lessgtr}(t)$ and the equations~(\ref{Sigma_time}) and~(\ref{eq:sigma_R})
\item Fourier transform the self-energy $\Sigma^R(t)$ to get $\Sigma^R(\omega)$
\item compute the new Green's function $G^{R} (\omega)$ from Dyson's equation~(\ref{eq:Dyson_eq})
\item if Green's function is converged, this is the GW approximation Green's function. If not, restart iterations at step 2.
\end{enumerate}

From the GW Green's function, the spectral matrix $A_{i\sigma,j\sigma'}$ is given by:
\begin{equation}
A_{i\sigma,j\sigma'} (\omega) = -2 \Im(G^R_{i\sigma,j\sigma'}(\omega)).
\label{eq:spectr_GR}
\end{equation}

The local density of states (LDOS, $n_{i\sigma}(\omega)$) is expressed using the diagonal elements of the spectral matrix and the total density of states (DOS, $D(\omega)$) with the sum of diagonal elements of the spectral matrix (\textit{i.e.} the trace):
\begin{equation}
n_{i\sigma}(\omega) = \frac{1}{2\pi} A_{i\sigma,i\sigma} (\omega)
\label{eq:ldos_GF}
\end{equation}
and
\begin{equation}
D(\omega) = \sum_{i\sigma} n_{i\sigma}(\omega) .
\label{eq:dos_GF}
\end{equation}

The matrix elements of the density matrix ($\textbf{n}$) are given by the integrals of the spectral function multiplied by the Fermi-Dirac statistics:
\begin{equation}
n_{i\sigma,j\sigma'} = \int_{-\infty}^{+\infty} \frac{\dd{\omega}}{2\pi} A_{i\sigma,j\sigma'} (\omega) f_{FD}(\omega)
\label{eq:dens_mat}
\end{equation}
and the local densities are the diagonal elements of the density matrix:
\begin{equation}
n_{i\sigma} = n_{i\sigma, i\sigma}.
\label{eq:density_expr}
\end{equation}

The total energy of the system is found using the Galitskii-Migdal formula~\cite{stefanucci_nonequilibrium_2013, joost_lowdins_nodate}:
\begin{equation}
E_{\rm Tot} = E_{\rm kin} + E_{\rm GM}
\label{eq:E_tot_GF}
\end{equation}
where $E_{\rm kin}$ is the kinetic energy and $ E_{\rm GM}$ is the interaction energy. They are given by:
\begin{equation}
\begin{split}
& E_{kin} = tr[\hat{H}_0 \textbf{n}] \\
& E_{GM} = -\frac{i}{2} \int \frac{\dd{\omega}}{2\pi} \omega \, tr[G^<(\omega)]
\end{split}
\label{eq:E_kin_GM_GF}
\end{equation}
where $\hat{H}_0$ is the non-interacting Hamiltonian, \textit{i.e.} the tight-binding part of Hubbard Hamiltonian or the first term of eq.~(\ref{Hubbard_ham}) proportional to the $t$ parameter.

The simulation of spatially-resolved $dI/dV$ images from STS (scanning tunneling spectroscopy) in the interval of energies $[E_1,E_2]$ can be obtained from the spectral matrix~\cite{joost_correlated_2019, meunier_tight-binding_1998}:
\begin{equation}
\frac{\dd{I}}{\dd{V}} (x,y,z_0) = \int_{E_1}^{E_2} \dd{\omega} \sum_{ij} A_{ij}(\omega) z_0^2 e^{\lambda^{-1} \abs{\vec{r}-\vec{r_i}}} e^{\lambda^{-1} \abs{\vec{r}-\vec{r_j}}}
\label{eq:stm_simu}
\end{equation}
where $z_0$ is the tip's height of the simulated $STS$ and $\lambda$ is a length parameter that parameterizes the spatial extension of localized orbitals. $(x,y,z_0)=\vec{r}$ is the location where the $STS$ is simulated and $\vec{r}_i$ are the atomic positions.

\subsection{Exact diagonalization} \label{sectionED}

In the exact diagonalization approach, we deal with the exact Hubbard Hamiltonian of eq.~(\ref{Hubbard_ham}). This implies a treatment in the many-electron basis rather than the simpler one-electron basis and, as a result,  the rapidly growing size of the Hilbert space. With only one orbital per atom, the size of the many-electron basis is given by $C_{N_{\rm el}}^{2N_{\rm at}} = \frac{2N_{\rm at}!}{(2N_{\rm at} -N_{\rm el} )! (N_{\rm el})!}$, where $N_{\rm at}$ is the number of atomic sites and $N_{\rm el}$ is the number of electrons in the system, in strong contrast to the linear scaling of the single-electron basis.

For example, the many-electron basis for a two-site Hubbard system containing two electrons includes six basis vectors that are given by $\ket{\Phi_1} = \ket{\uparrow, \uparrow} , \ket{\Phi_2} = \ket{\uparrow, \downarrow}, \ket{\Phi_3} = \ket{\downarrow, \uparrow}, \ket{\Phi_4}  = \ket{\uparrow \downarrow, .}, \ket{\Phi_5} \ket{., \uparrow \downarrow} $ and $\ket{\Phi_6} =  \ket{\downarrow, \downarrow}$ where the comas symbolically separate atomic sites, the dots represent empty sites, and $\uparrow$ (resp. $\downarrow$) stands for the presence of a spin-up (resp. down) electron.

We follow the method outlined in Refs.~\cite{lin_exact_1993, jafari_introduction_2008,kingsley_exact_2013, sharma_organization_2015}. It consists in labelling the state with one integer $I$, bijectively linked to two other integers $I_{\uparrow}$ and $I_{\downarrow}$ by the relations:
\begin{equation}
I=2^{N_{\rm at}} I_{\uparrow} + I_{\downarrow}
\label{eq:I_tot}
\end{equation}
and
\begin{equation}
\begin{split}
&I_{\uparrow} = I//2^{N_{\rm at}} \\
&I_{\downarrow} = I \mod{2^{N_{\rm at}}} 
\end{split}
\label{eq:relation_I_Iup_Idown}
\end{equation}
where $//$ represents the integer division.

Writing $I_{\uparrow}$ (resp. $I_{\downarrow}$) in binary notation gives the space configuration of the states in the spin up (resp. down) sector. Organizing the Hilbert space in this way allows one to treat only integers and to easily find the actions of creation, destruction, and density operators on each state using simple standard binary operations (bin flip, bin counting,\ldots). We explicitly show the correspondence between basis states and $I$, $I_\uparrow$ and $I_\downarrow$ for the two-site Hubbard system at half-filling in table~\ref{tab:I_Iup_Idown_two_site}.

\begin{table}
    \centering
    \begin{tabular}{c|c|c|c|c}
        Basis state  & $I_\uparrow$ & $I_\downarrow$  & $I$ \\
        \hline
        $\ket{\uparrow, \uparrow}$  & 3 & 0  & 12 \\
        $\ket{\uparrow, \downarrow}$ & 2 & 1 & 9  \\
        $\ket{\downarrow, \uparrow}$  & 1 & 2 & 6 \\
        $\ket{\uparrow \downarrow, .}$  & 2 & 2 & 10 \\
        $\ket{., \uparrow \downarrow}$  & 1 & 1 & 5 \\
        $\ket{\downarrow, \downarrow}$  & 0 & 3 & 3
    \end{tabular}
    \caption{Illustration of the indexing scheme used to build many-electron basis states for the two-site Hubbard system at half-filling. $I_\uparrow$ follows from the binary notation of spin up occupation of the states from right to left: the first state for example has one spin up electron on site $0$ (right) and one spin up electron on site $1$ (left) such that $I_{\uparrow}=2^0 + 2^1 = 3$. The same reasoning for spin down electrons leads to determination of $I_{\downarrow}$ and $I$ follows from the formula~(\ref{eq:I_tot}).}
    \label{tab:I_Iup_Idown_two_site}
\end{table}

In the ED framework, the retarded Green's function describing the system with $N_{\rm el}$ electrons involves the Hamiltonian operators acting in the Hilbert space containing states representing $N_{\rm el}+1$ and $N_{\rm el}-1$ electrons ($\hat{H}_{N_{el}+1}$ and $\hat{H}_{N_{el}-1}$). Its definition for a general system of $N_{\rm el}$ electrons in the frequency domain is~\cite{negele_quantum_1988}:
\begin{equation}
\begin{split}
G^R_{i\sigma, j\sigma'} (\omega) =& \\ &\mkern-40mu\bra{\Psi_0^{N_{el}}} \hat{c}_{i\sigma} \frac{1} {\omega + (E_0^{N_{el}}-\hat{H}_{N_{el}+1} +i\eta)} \hat{c}^\dagger_{j\sigma'} \ket{\Psi_0^{N_{el}}}+ \\
& \mkern-40mu \bra{\Psi_0^{N_{el}}} \hat{c}^\dagger_{i\sigma} \frac{1} {\omega - (E_0^{N_{el}}-\hat{H}_{N_{el}-1}-i\eta)} \hat{c}_{j\sigma'} \ket{\Psi_0^{N_{el}}}
\end{split}
\label{eq:many_body_greens_function}
\end{equation}
where $\ket{\Psi_0^{N_{el}}}$ and $E_0^{N_{el}}$ are the ground state and the ground state energy of the system with $N_{el}$ electrons. $\eta$ is a small real positive parameter, introduced to ensure causality.

The first term of eq.~(\ref{eq:many_body_greens_function}) is the electron addition part of the Green's function.  $\hat{c}^\dagger_{j\sigma'} \ket{\Psi_0^{N_{el}}}$ and $\bra{\Psi_0^{N_{\rm el}}} \hat{c}_{i\sigma}$ represent both states with $N_{\rm el}+1$ electrons. It involves the matrix elements between these states and the $N_{\rm el}+1$ electrons Hamiltonian (at the denominator). This term thus explores the possible states when an electron is added from the $N_{\rm el}$ ground state. In the second term of eq.~(\ref{eq:many_body_greens_function}), one electron is annihilated instead of created, this is thus the electron removal part of the Green's function.

In order to evaluate the retarded Green's function~(\ref{eq:many_body_greens_function}), three exact diagonalizations are required: the first one is done in the $N_{\rm el}$ sector to find the ground state $\ket{\Psi_0^{N_{el}}}$ and its energy $E_0^{N_{el}}$. Then an ED is completed in the sectors $N_{el}\pm1$ so that matrix elements of the type $\bra{\Phi_k^{N_{el}\pm1}} \hat{H}_{N_{el}\pm1} \ket{\Phi_{k'}^{N_{el}\pm1}}$ can be computed. $\ket{\Phi_{k'}^{N_{el}\pm1}}$ are basis vectors of the $N_{el}\pm1$ sector of the Hilbert space.

The retarded Green's function is then computed by first expressing the states $\hat{c}^\dagger_{j\sigma'} \ket{\Psi_0^{N_{el}}}$ in the basis $\ket{\Phi_{k}^{N_{el}+1}}$ and the states $\hat{c}_{j\sigma'} \ket{\Psi_0^{N_{el}}}$ in the basis $\ket{\Phi_{k}^{N_{el}-1}}$. 

From the evaluation of the retarded Green's function~(\ref{eq:many_body_greens_function}), physical quantities such as the LDOS, DOS, density matrix, and the $dI/dV$ STS simulated maps can be computed (eqs.~(\ref{eq:ldos_GF}),~(\ref{eq:dos_GF}),~(\ref{eq:dens_mat}), and~(\ref{eq:stm_simu})). The total energy of the system, $E_0^{N_{\rm el}}$, is obtained using a single ED.

\begin{figure}
\centering
    \includegraphics[width=.6\textwidth]{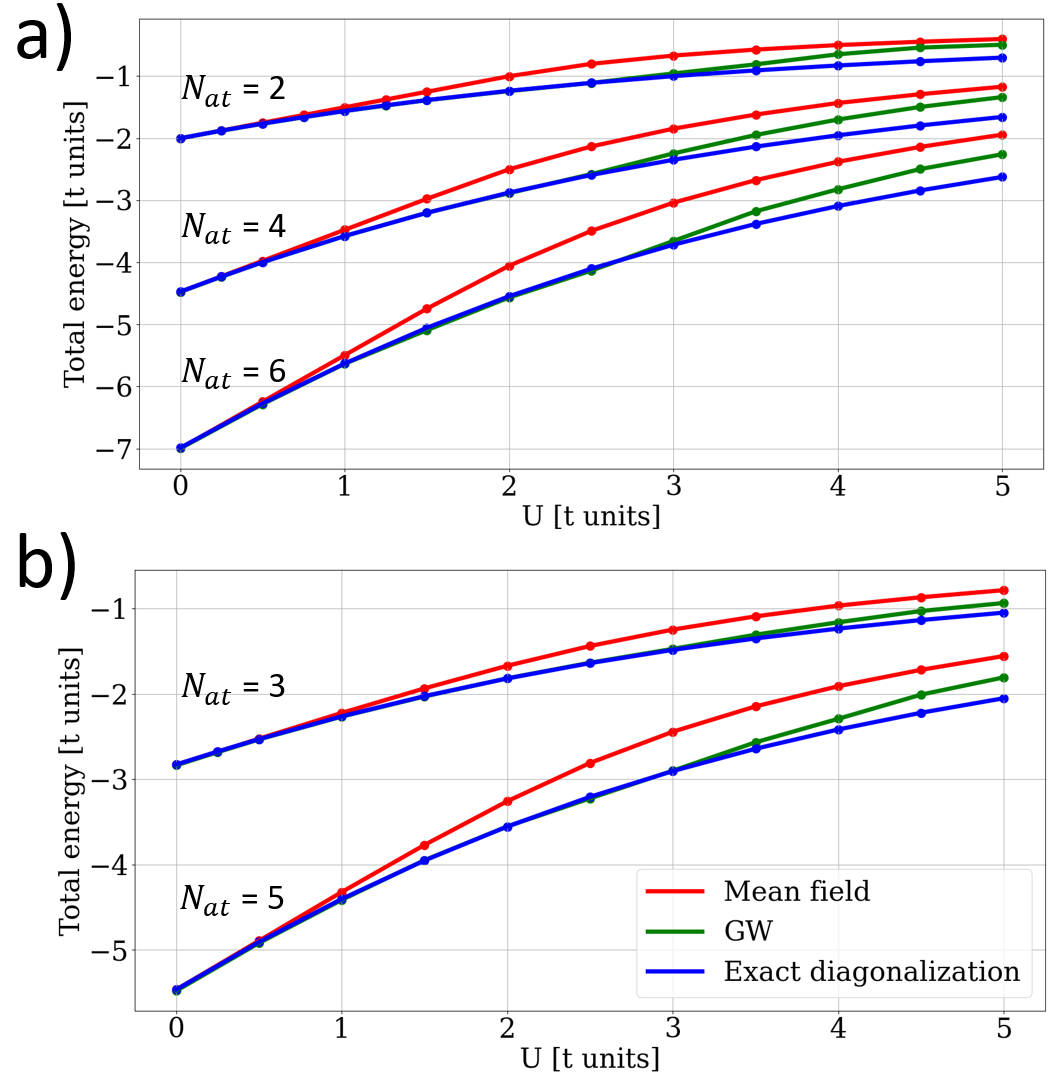}
\caption{Total energy at half-filling as a function of the $U$ parameter of the Hubbard model in linear chains with even number of sites (a) and odd number of sites (b). Results are given for MF (in red), GW (in green), and ED (in blue).}
\label{fig:Etot_all_lengths_2}
\end{figure}

\subsection{Magnetic order parameters}

We have already mentioned that the magnetic symmetry (or order) requires a good description of the correlation. We define here two order parameters that will allow us to characterize the phase transition between the PM, FM and AFM phases. The local magnetic moments ($M_i$) are defined as the difference between spin-up and spin-down local mean densities and the total spin ($S$) is the sum of the local magnetic moments~\cite{yazyev_emergence_2010}:
\begin{equation}
M_i = \frac{n_{i\uparrow}  - n_{i\downarrow} }{2}
\end{equation}
and 
\begin{equation}
S = \sum_i M_i.
\label{eq:total_spin}
\end{equation}

Finally, the staggered magnetization ($M$) is defined as the sum of absolute values of the magnetic moments:
\begin{equation}
M= \sum_i \abs{M_i}.
\label{eq:stag_magn}
\end{equation}

A vanishing $M$ corresponds to the spin-symmetric case and thus to the $PM$ phase since $M_i=0$ for all $i$. When $M$ is non-zero, the state is not spin-symmetric and is thus either $FM$ or $AFM$. In the case of even number of electrons, $S$ helps identifying the $FM$ and $AFM$ phases since it is non-zero for $FM$ states and zero in the $AFM$ case.

\section{Global properties of the systems}
\label{sec:global_prop}

We first study the total energy of linear chains of varying lengths (from $N_{\rm at}=2$ to $N_{\rm at}=6$) with open boundary conditions. All chains are studied at half electronic filling, \textit{i.e.} with the number of electrons equal to the number of sites. The important parameter of the Hamiltonians is the ratio $U/t$ and, in this paper, all results are expressed in units of $t$.

We distinguish the cases of even and odd number of atoms in the chain because the total spin (see eq.~(\ref{eq:total_spin})) of the ground state is predicted to be either $0$ or $1/2$ for even and odd cases respectively, in agreement with Lieb's theorem~\cite{lieb_two_1989, yazyev_emergence_2010}. In the even case, this implies that the ground state cannot be FM but must be PM or AFM. In the odd case, the PM phase is forbidden due to Lieb's theorem.

Total energies computed by the three methods (MF, GW, and ED) as a function of the $U$ parameter are shown in figure~\ref{fig:Etot_all_lengths_2}. For even numbers of atoms, we obtain results in agreement with those reported in  Ref.~\onlinecite{joost_lowdins_nodate} for both MF and ED approaches (Ref~\onlinecite{joost_lowdins_nodate} does not report results for odd number of atoms). Random initial occupations have been used in the case of the MF approximation such that it has to be compared with unrestricted-spin Hartree-Fock ("usHF") method employed in Ref.~\onlinecite{joost_lowdins_nodate}. For small values of $U$, MF and GW are in good agreement with the exact ED and display a linear increase of the total energy with U ~\cite{joost_lowdins_nodate}. GW approximation lowers the total energy from MF and agrees very well with ED until $U=3t$, thereby doubling the range of very good agreement compared to MF.

\begin{figure*}
\centering
    \includegraphics[width=\textwidth]{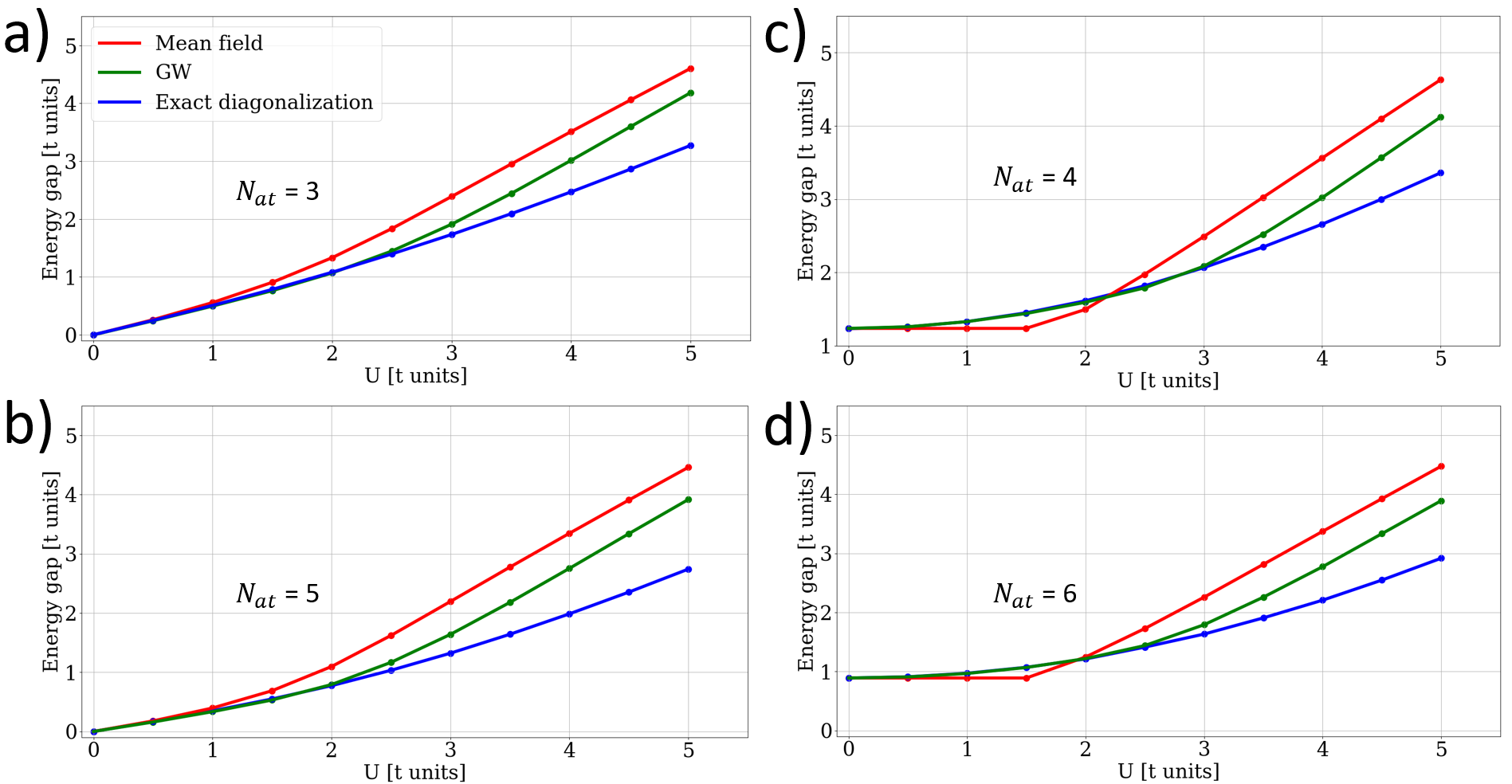}
\caption{Evolution of the energy gap at half-filling as a function of the $U$ parameter of the Hubbard model in linear chains with odd number of sites (a and b) and even number of sites (c and d). Results are given for MF (in red), GW (in green), and ED (in blue).}
\label{fig:EG_Nat_3_4_5_6}
\end{figure*}

The energy gap (energy interval between the highest peak in the DOS below the Fermi level and the lowest peak above the Fermi level) of the same linear chains are shown in figure~\ref{fig:EG_Nat_3_4_5_6}. For an odd number of sites (Fig.~\ref{fig:EG_Nat_3_4_5_6} (a) and (b))), the energy gap is zero at $U=0$ and grows with increasing $U$. This growth is overestimated by MF. MF deviates significantly from ED above $U \simeq t$ where the GW approximation follows the ED curve more closely until $U\simeq 2t-2.5t$. Then it deviates in turn with ED, lowering the energy gap from MF but sill overestimating it when compared with ED. For even number of atoms ((Fig.~\ref{fig:EG_Nat_3_4_5_6} (c) and (d))), the energy gap is non-zero for $U=0$. It increases with $U$ for ED but remains constant until $U=1.5t$ and starts increasing around $U=2t$ in MF. As shown in~\cite{joost_lowdins_nodate}, this sudden increase in MF is due to a phase transition as the ground state for small values of $U$ is PM whereas it is AFM for larger values. This transition will be discussed later on considering the staggered magnetization as an order parameter. According to Ref.~\onlinecite{joost_lowdins_nodate}, the increase of energy gap in ED is due to an increase in correlation, that is absent in MF. In MF, the correlation is mimicked by a symmetry breaking of the ground state from a homogeneous spin-symmetric state (PM) into a AFM state, increasing suddenly the energy gap. However, the actual effect of correlation is not included satisfactorily as ED remains PM. As GW is based on the MF solution, chosen as a starting point, it converges towards a state with the same symmetry. It follows that for small $U$, GW results in figures~\ref{fig:Etot_all_lengths_2} and~\ref{fig:EG_Nat_3_4_5_6} are for a PM state whereas it is for an AFM at larger $U$. For $U\le 1.5t$, the increase in the energy gap for even number of atoms chains (see figure~\ref{fig:EG_Nat_3_4_5_6} c) and d)) can thus presumably be attributed to correlation effects, absent in MF. Note that in odd number of atoms chains, there is no phase transition such that all three models predict AFM for all values of $U$. Still, there is also an improvement in the energy gap with the GW correction.

\begin{figure}
\centering
    \includegraphics[width=.8\textwidth]{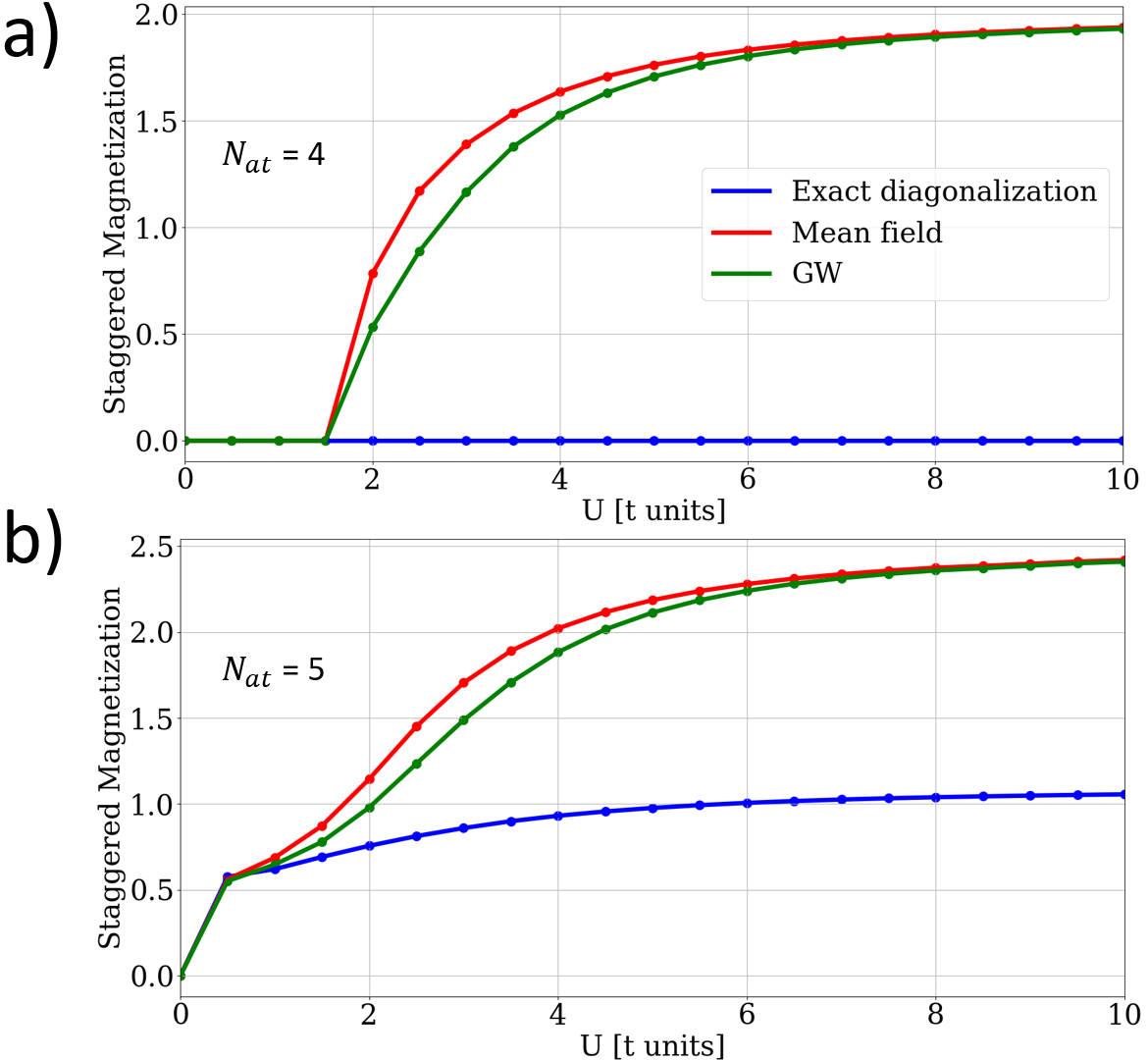}
\caption{Staggered magnetization of the linear chains for $N_{at}=4$ (a) and for $N_{at}=5$ (b). Results are given for MF (in red), GW (in green), and ED (in blue). The zero values in a) indicates a PM phase (in all regimes for ED and for $U\le t$ in MF and GW). In other cases, the state is AFM.}
\label{fig:stag_magn}
\end{figure}

 $M$ is the order parameter distinguishing the PM and the AFM phases and we analyze it here in details for the even-number-of-atoms case. The staggered magnetization as a function of the $U$ parameter is given in figure~\ref{fig:stag_magn} for $N_{\rm at}=4$ and $5$ to illustrate the even and odd cases respectively. There is globally three different regimes in terms of the $U$ parameter: very small $U$ ($U\le 1.5t$), intermediate $U$ ($1.5t \le \hspace{.1cm} U \le 6t$), and large $U$ ($U\ge 6t$). In the odd number of atoms case at small $U$, $M$ increases, starting from $M=0$, in the three approaches. In the intermediate regime, MF overestimates $M$ and GW tends to lower it in better agreement with ED. In the large $U$ regime, the MF and GW staggered magnetization $M$ have almost the same values that are much larger than for ED. In the case of even number of atoms, $M$ is zero for small $U$ for all methods, indicating a PM phase. The PM phase is maintained for all $U$ in ED. However, a phase transition occurs towards an AFM phase in MF and GW resulting in a non-zero staggered magnetization. In the intermediate regime, the GW corrects the MF approximation in the right direction (that is, it decreases $M$) where GW do not correct effectively MF anymore for larger $U$.

To summarize, the GW approximation reproduces correlations well in the small $U$ regime for both even and odd cases, it tends to reduce the overestimated staggered magnetization of MF in the intermediate regime and finally barely corrects this quantity in the large $U$ regime. 
We note however that the GW approximation results considered in this section exhibit the same phase transition as in MF approximation since the latter is the starting point of former. This phase transition is not present in the ED and is not directly visible when looking at the total energies. We discuss the change in the phase transition in a following  section.

\begin{figure}
\centering
    \includegraphics[width=.6\textwidth]{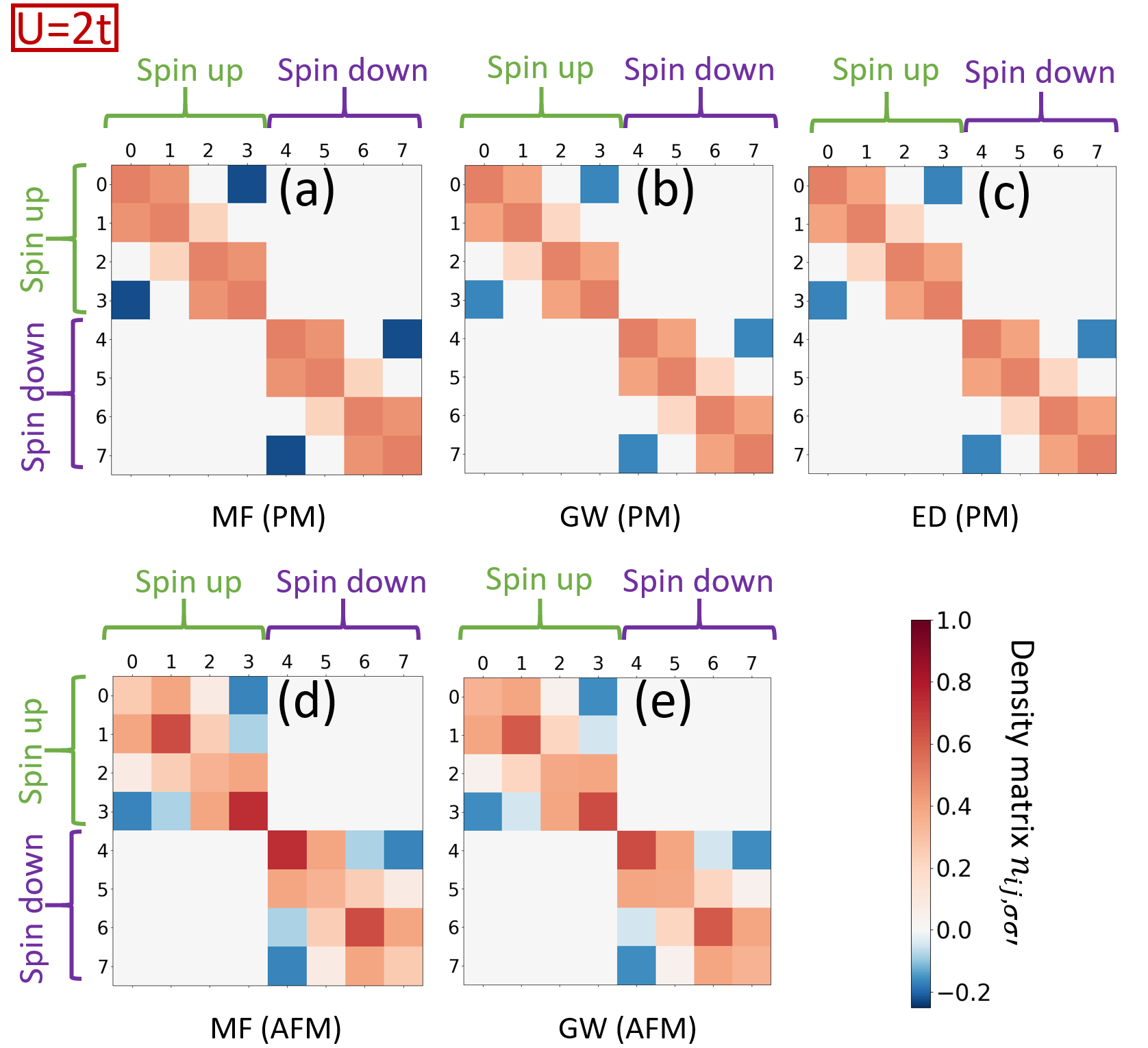}
\caption{Representation of the density matrices for a $N_{at}=4$ atomic chain at half-filling and at $U=2t$. The density matrix are given for the PM states in MF (a), GW (b) and ED (c) and for the AFM states in MF (d) and GW (e). According to the colorbar, positive values are in red and negative values are in blue. Labels from $0$ to $3$ (resp. from $4$ to $7$) represent spin up (resp. down) electron on atomic sites from $0$ to $3$.}
\label{fig:dens_mat_Nat_4_U_2t}
\end{figure}

\begin{figure}
\centering
    \includegraphics[width=.6\textwidth]{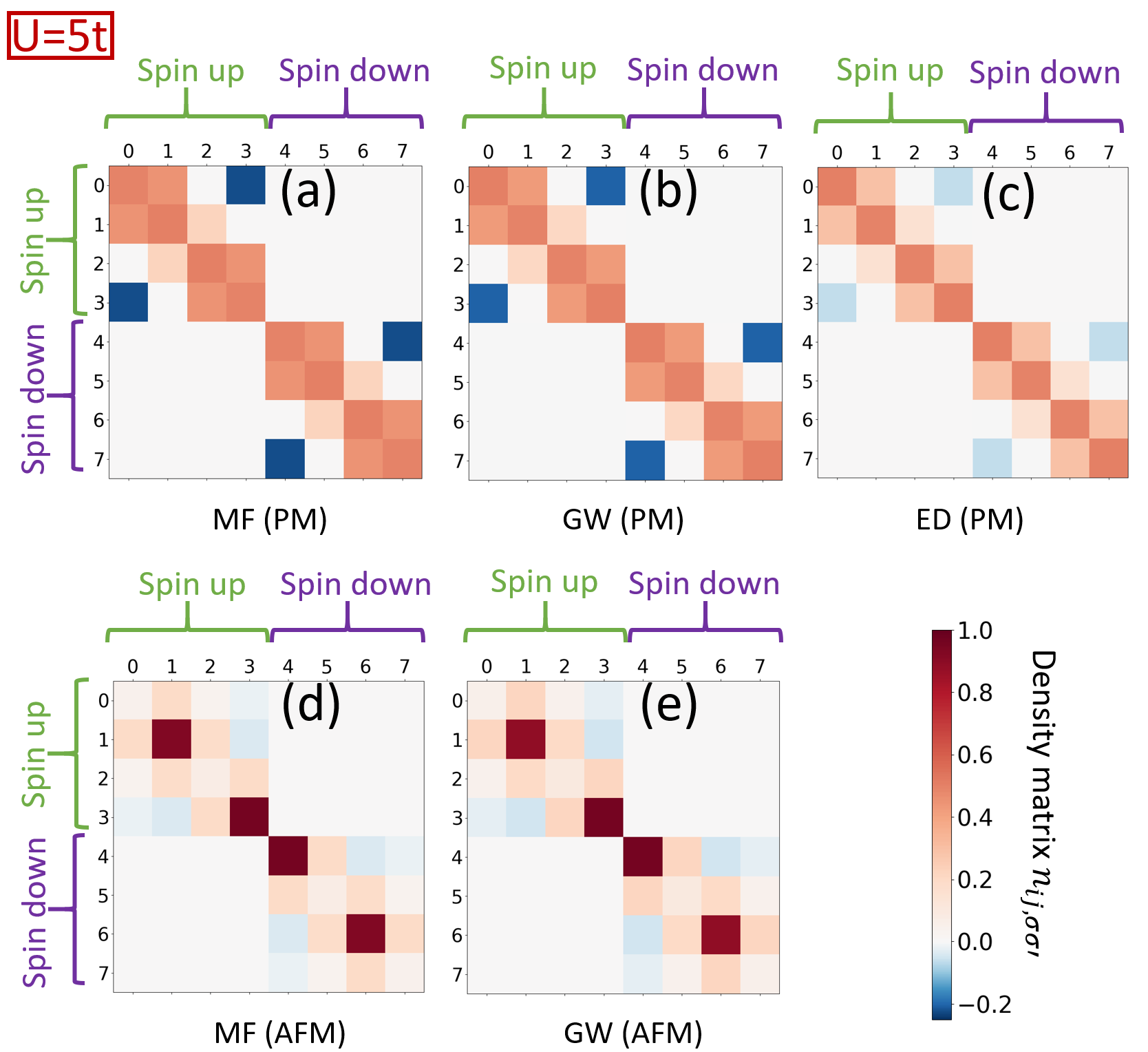}
\caption{Same as figure~\ref{fig:dens_mat_Nat_4_U_2t} with $U=5t$.}
\label{fig:dens_mat_Nat_4_U_5t}
\end{figure}

The analysis of density matrices allows to compare local densities of the different states as well as non-diagonal coupling between atomic sites. The density matrices (see eq.~(\ref{eq:dens_mat})) for ED and the PM and AFM states of MF and GW approximations are represented graphically in figure~\ref{fig:dens_mat_Nat_4_U_2t} for $U=2t$ and in figure~\ref{fig:dens_mat_Nat_4_U_5t} for $U=5t$. The diagonal elements are local electronic densities such that a uniform diagonal corresponds to a PM state whereas an AFM is characterized by diagonal elements that exhibit high values for spin up and spin down electrons, alternatively. We first observe that the PM states computed in MF and GW have a similar appearance compared to those obtained by ED (see \textit{e.g.} figure~\ref{fig:dens_mat_Nat_4_U_2t} a), b) and c)). This is not the case for the AFM states (compare for example figure~\ref{fig:dens_mat_Nat_4_U_2t} d) and e) with c)). 

For $U=2t$, the density matrix of the PM state in MF (figure~\ref{fig:dens_mat_Nat_4_U_2t} (a)) presents a strong negative coupling between sites $0$ and sites $3$, \textit{i.e.} the edges of the chain. This coupling is reduced in GW (figure~\ref{fig:dens_mat_Nat_4_U_2t} (b)), in better agreement with ED (figure~\ref{fig:dens_mat_Nat_4_U_2t} (c)). Then, the coupling between neighboring sites (site $0$ - site $1$ and site $2$ - site $3$) is also reduced in absolute value from MF to GW, still better reproducing the ED result. Considering AFM states (figures~\ref{fig:dens_mat_Nat_4_U_2t} (d) and (e)), the anti-ferromagnetic character is decreased between MF and GW as the contrast between alternating diagonal elements is decreased. The negative couplings between sites $1$ and $3$ for spin up and between sites $0$ and $2$ for spin down are reduced from MF to GW, but is still non-zero in GW contrary to ED.

Increasing the $U$ parameter to $U=5t$ (see figure~\ref{fig:dens_mat_Nat_4_U_5t}) leads to a similar GW correction: the negative coupling between sites $0$ and $3$ and the positive coupling between adjacent sites are reduced. However, even if GW leads to a more accurate description of density matrices, the differences with ED density matrix are greater than for $U=2t$. This is an expected feature since the GW approximation is a perturbative method in terms of the $U$ parameter. For the AFM states, the alternating behavior in the diagonal of the density matrix is also reduced from MF to GW, and some other minor changes occur in non-diagonal elements but both MF and GW density matrices are far away from the ED one, that shows a PM state. The correction from MF to GW is less important than for $U=2t$.

\begin{figure}
\centering
    \includegraphics[width=.8\textwidth]{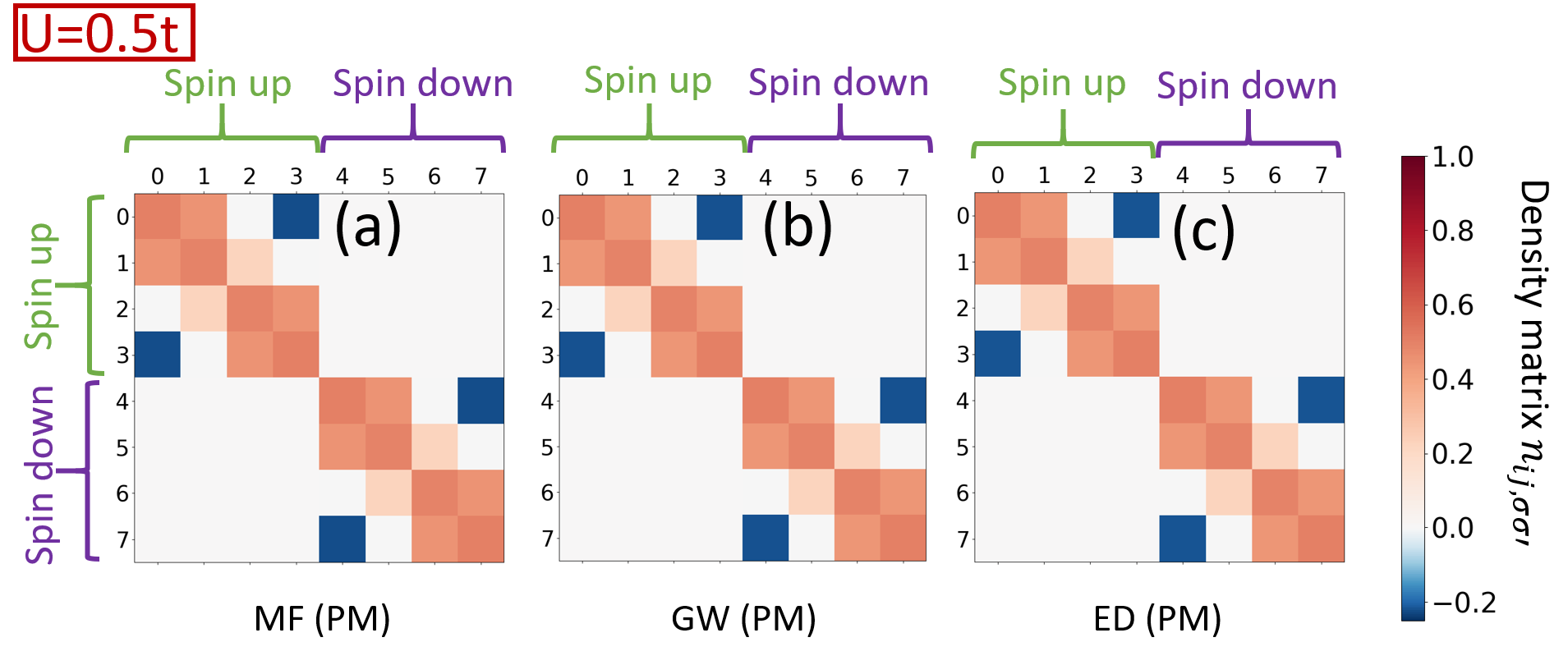}
\caption{Same as figure~\ref{fig:dens_mat_Nat_4_U_2t} with $U=0.5t$ and only for PM states.}
\label{fig:dens_mat_Nat_4_U_0p5t}
\end{figure}

To gain a more precise view of the evolution of the density matrices as a function of $U$, we also report density matrices in the case of small $U$ ($U=0.5t$) in figure~\ref{fig:dens_mat_Nat_4_U_0p5t}. As the value is below the transition, there are no AFM states in MF (nor in GW). MF and GW are here good approximations and lead to very similar density matrices as those obtained by ED. More interesting is the evolution of density matrices of PM states within the same approximation as a function of $U$, comparing figs.~\ref{fig:dens_mat_Nat_4_U_2t} (a) (resp. (b) and (c)),~\ref{fig:dens_mat_Nat_4_U_5t} (a) (resp. (b) and (c)) and~\ref{fig:dens_mat_Nat_4_U_0p5t} (a) (resp. (b) and (c)) for the MF (resp. GW and ED) case. In the MF case, very small changes are observed when increasing the $U$ parameter. In the ED case, negative couplings between sites $0$ and $3$ as well as positive coupling between sites $0$ and $1$ and between sites $2$ and $3$ are reduced in absolute values. The interesting point is that the GW approximation exhibits the same trend as ED from small to intermediate values of $U$, but the opposite one when going to large $U$. Indeed, from $U=2t$ to $U=5t$, coupling strengths increase in GW as mentioned above (after having decreased from $U=0.5t$ to $U=2t$), where they keep on decreasing in ED.

\section{DOS, LDOS and simulated $STS$ maps}
\label{sec:freq_dep}

The previous section focused on global properties such as total energy, total staggered magnetization, energy gap, and density matrix. We now turn to frequency-dependent properties. These quantities are also of interest as they can be interpreted as single-particle features (energy, density,\ldots) in the MF approximation and can thus offer further insight into the effects of many-body interactions and the missing correlation in MF by comparison with ED results. They are also measurable properties or can be used to compute measurable quantities (\textit{e.g.}, $STS$ maps, optical absorption, plasmons,\ldots~\cite{joost_correlated_2019, honet_semi-empirical_2021})

We start by investigating the effect of GW on a two-site system so that we can make the connection with previously published analytical studies~\cite{di_sabatino_reduced_2015, romaniello_self-energy_2009, romaniello_beyond_2012} and then extend our study to larger systems as those in previous sections. We note that in the cited analytical studies, the so-called "GW approximation" is actually the "$G_0W_0$ approximation" in the sense that no self-consistency procedure was introduced and the GW results are identified to the results obtained after the first step of computation. 

\begin{figure}
\centering
    \includegraphics[width=.8\textwidth]{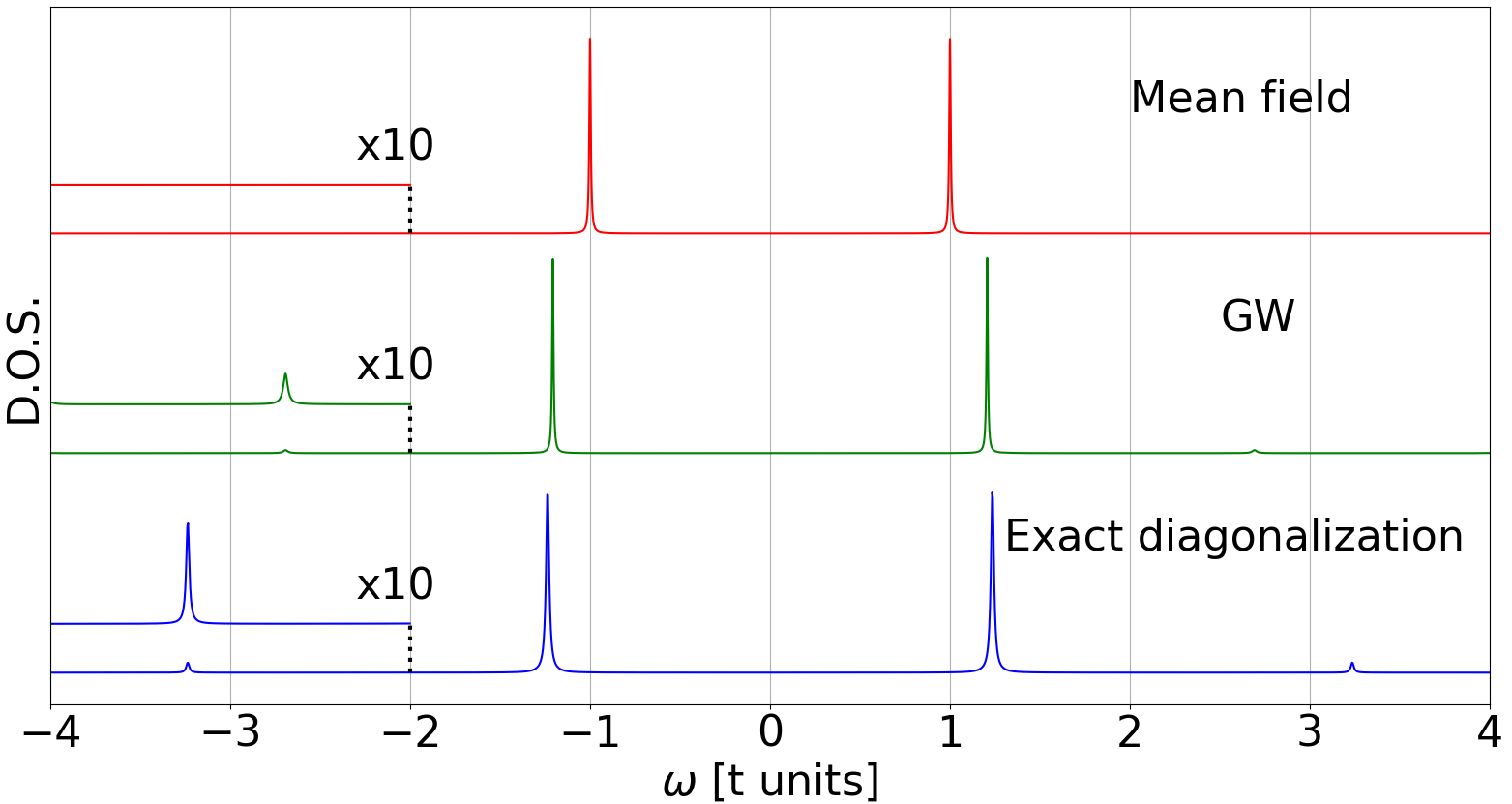}
\caption{DOS of the half-filled for $N_{at}=2$ Hubbard chain at $U=2t$ for the MF approximation (in red), GW approximation (in green) and ED (in blue). All Fermi energies have been aligned to $0$. The small panel above the principal graph shows a zoom of the DOS between $U=-4t$ and $U=-2t$ to better distinguish the satellites.}
\label{fig:DOS_two_site}
\end{figure}

\begin{figure}
\centering
    \includegraphics[width=.8\textwidth]{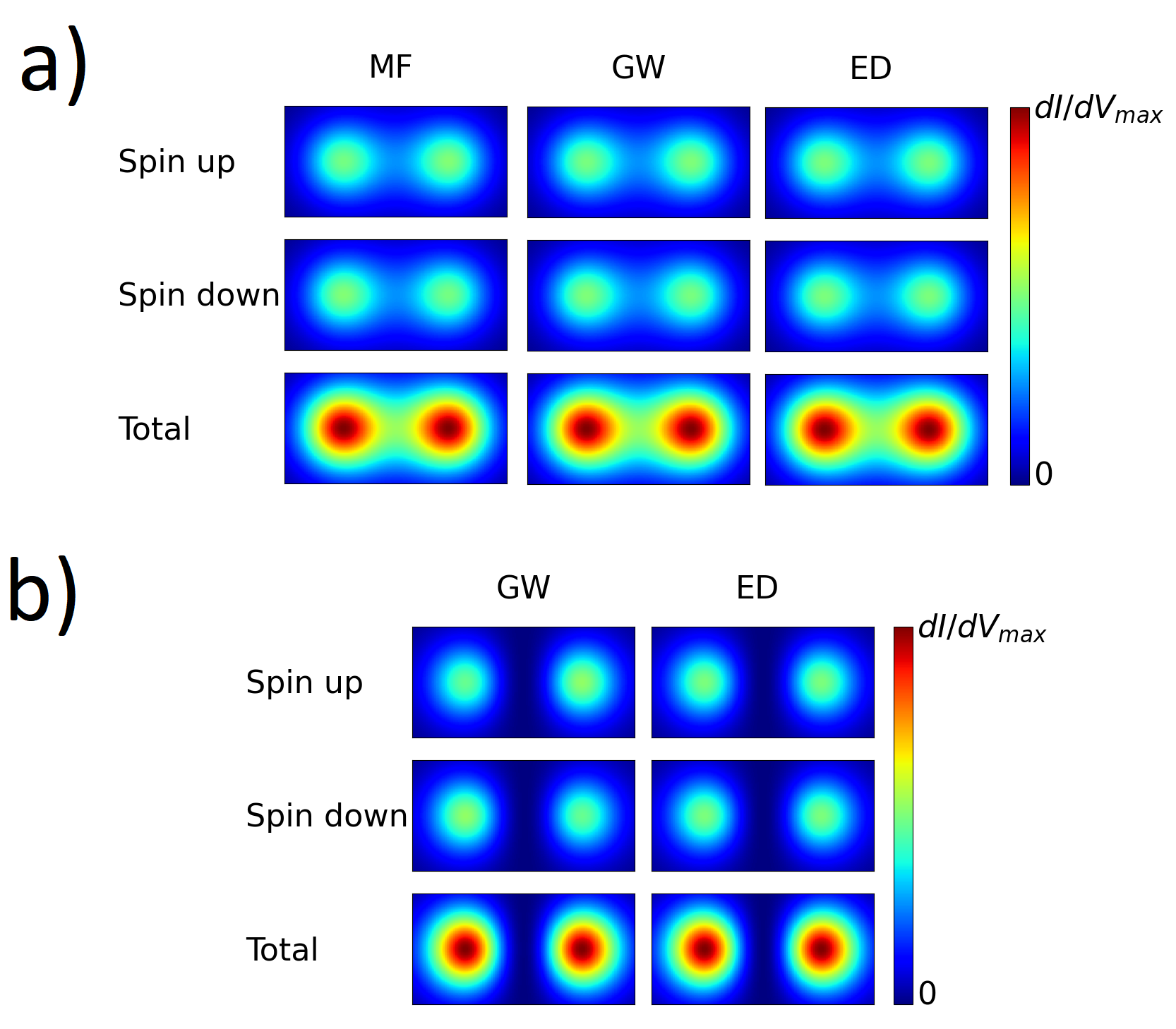}
\caption{Simulated STS maps associated with the main DOS peak below the Fermi level (a) and for the principal satellite feature below the Fermi level (b) for the two-site half-filled Hubbard system at $U=2t$. For the main peak (a), MF approximation is at top left, GW approximation is at top right and ED is at the bottom. For the satellite (b), GW approximation is at the left and ED at the right (there is no satellite for MF). For each method, the spin up and spin down channels are shown separately and the sum of the channels is labelled "total". The intensities have been independently normalized. The size of the scanning area is $2 d_{i-a} \cross d_{i-a}$ in each case with $d_{i-a}$ the inter-atomic distance.}
\label{fig:dIdV_two_site_QP_sat}
\end{figure}

All results shown in this section are obtained at an intermediate value of the Hubbard parameter ($U=2t$) and all the Fermi levels have been aligned to $0$ in the study of DOS and LDOS for better visualization. The DOS for the two-site system in MF, GW, and ED are shown in figure~\ref{fig:DOS_two_site}. The DOS in MF consists only in two doubly degenerated peaks at $\omega=\pm t$ that can be attributed to single-particle states with bonding (resp. anti-bonding) symmetry for the state below (resp. above) the Fermi level. Bonding states (resp. anti-bonding states) are characterized by a non-zero (resp. zero) intensity at the mid-point between the atoms. This is consistent with ED results. In GW, these main peaks are shifted away from the Fermi level at $\omega \simeq \pm 1.21 t $, increasing the gap and resulting in a value closer to the result of ED ($\omega \simeq \pm 1.24 t$). However, these main peaks do not integrate to $1$ in GW and ED and cannot be attributed to single-particle states but to quasi-particle (QP) peaks, as explained in Ref.~\onlinecite{reining_gw_2018}. The symmetry of the QP states is preserverd and corresponds to ED (see figure~\ref{fig:dIdV_two_site_QP_sat} (a) for the QP peak below the Fermi level). 

A closer look at the GW approximation DOS reveals smaller satellite peaks at $\omega \simeq \pm 2.69 t$ for GW and  $\omega \simeq \pm 3.24 t $ for ED, surrounding the QP ones~\cite{reining_gw_2018}. This is another evidence of the improvement from MF to GW although the peaks are not exactly at the correct energies. However, the GW approximation predicts many satellites (of very small intensities) whereas only one on each side of the Fermi level is obtained in ED. As shown analytically in Ref.~\onlinecite{tomczak_proprietes_2007}, the satellite below (resp. above) the Fermi level have an anti-bonding (resp. bonding) character, that is the opposite symmetry than the nearest QP peak. This feature is reproduced well by the principal satellites introduced by GW as shown in figure~\ref{fig:dIdV_two_site_QP_sat} (b).

\begin{figure*}
\centering
    \includegraphics[width=\textwidth]{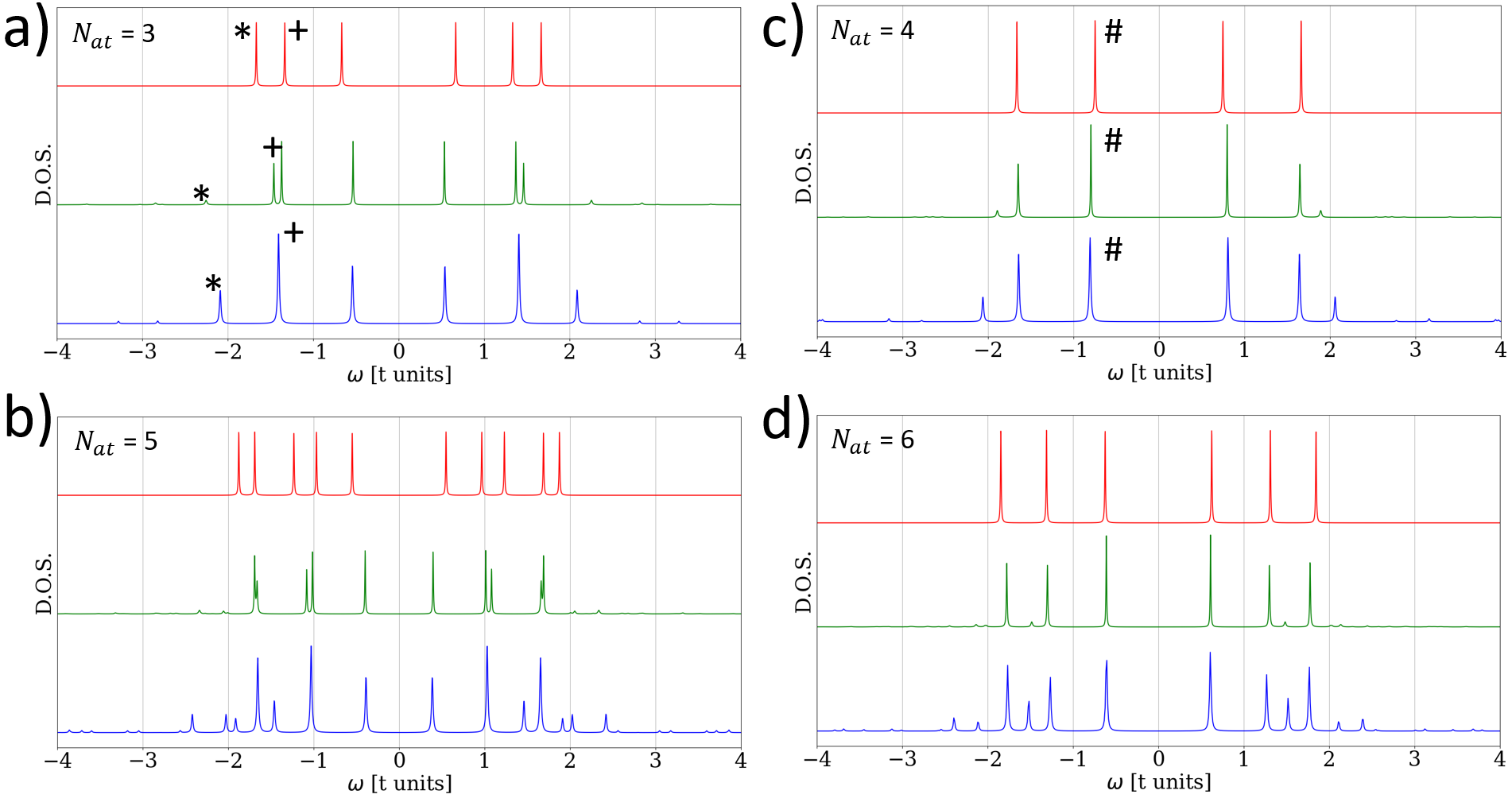}
\caption{DOS of half-filled Hubbard chains of different lengths ($N_{at}=3$ in (a), $N_{at}=5$ in (b), $N_{at}=4$ in (c) and $N_{at}=6$ in (d)). Red curves correspond to MF approximation, green curves to GW approximation and blue curves to ED. All Fermi energies have been aligned to $0$ for better visualization. +, * and $\#$ symbols are used to mark the peaks as references for figure~\ref{fig:Nat_3_sat_double_Nat_4_LUMO}.}
\label{fig:DOS_Nat_3_4_5_6}
\end{figure*}

The DOS for $N_{at}=3,~4,~5$ and $6$ chains are shown in figure~\ref{fig:DOS_Nat_3_4_5_6}. As previously in the two-site system, we observe the energy correction of main peaks and the appearance of satellites in GW, in agreement with ED. For example, we see a clear energy shift for the MF peaks at $E\simeq \pm 0.67$ for $N_{at}=3$ and at $E\simeq \pm 0.55 $ for $N_{at}=5$. Splitting of the MF DOS peak in the GW approximation are observed, for example for the peaks at $E \simeq \pm 1.67 t$ for $N_{at}=3$, $E \simeq \pm 1.66 t$ for $N_{at}=4$, $E \simeq \pm 1.46 t$ and $E \simeq \pm 1.88 t$ for $N_{at}=5$ and $E \simeq \pm 1.31 t$ and $E \simeq \pm 1.85 t$ for $N_{at}=6$. Satellites are clearly identified in figures~\ref{fig:DOS_Nat_3_4_5_6} a) and b) at the outermost part of the spectrum after the main peaks at each side of the Fermi level in GW but also in between QP peaks as in figure~\ref{fig:DOS_Nat_3_4_5_6} d) at $E\simeq \pm 1.39t$ in GW. Interestingly, the shifting and splitting of MF states also result in bringing some peaks closer to each other as for example for the GW peaks of figure~\ref{fig:DOS_Nat_3_4_5_6} a) at $E\simeq \pm 1.37t$ and $E\simeq \pm 1.46t$. 

\begin{figure*}
\centering
    \includegraphics[width=\textwidth]{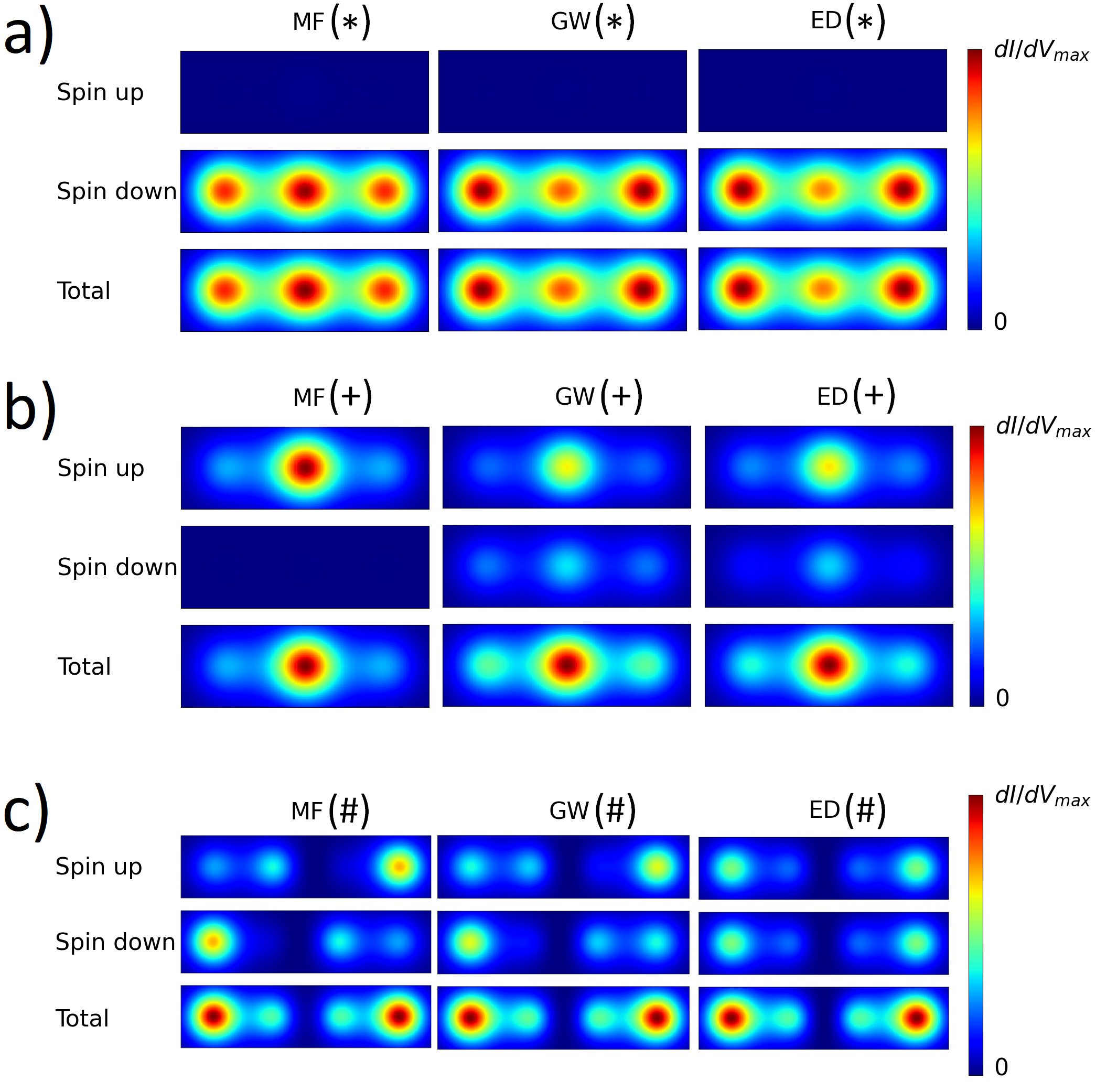}
\caption{(a) Simulated STS maps for the particle peak of MF at $E \simeq -1.66t$, the satellite at $E\simeq -2.25t$ for GW and the peak at $E\simeq  -2.09t$ for ED, for the $N_{at}=3$ Hubbard chain at half-filling. (b) Simulated STS maps for the particle peak of MF at $E \simeq -1.33t$, the two close peaks at $E\simeq  -1.37t$ and $E\simeq  -1.46t$ for GW and the large peak at $E\simeq  -1.41t$ for ED, for the $N_{at}=3$ Hubbard chain at half-filling.  (c) Simulated STS maps for the QP peaks just bellow the Fermi level, for the $N_{at}=4$ Hubbard chain at half-filling. MF peak is at $E \simeq -0.75t$, GW peak is at $E\simeq -0.80t$ and the ED peak at $E\simeq  -0.81t$. For each method and each subplot, the spin up and spin down channels are shown separately and the sum of the channels is labelled "Total". The intensities have been independently normalized. The size of the scanning area is $(N_{at} * d_{i-a}) \cross d_{i-a}$ in each case with $d_{i-a}$ the inter-atomic distance. +, * and $\#$ symbols are used as references for peak identification (see figure~\ref{fig:DOS_Nat_3_4_5_6}).}
\label{fig:Nat_3_sat_double_Nat_4_LUMO}
\end{figure*}

We now focus on the simulated $STS$ maps for $N_{at} = 3$. The one to one comparison of the STS maps is needed to correctly assign the DOS peaks to states with a given symmetry for each approaches. The lower state in MF ($E \simeq 1.67$, see the $*$ symbol) is compared to less intense peaks below $2t$ in GW and ED (see fig.~\ref{fig:Nat_3_sat_double_Nat_4_LUMO} a)). For the second lower energy state in MF, the GW approximation predicts two near peaks that better reproduce the STS map of ED greater peak (at $E\simeq \pm 1.41t$) when integrated together (see the $+$ symbol). The initial state in MF at $E\simeq \pm 1.67t$ thus splits into a satellite that is further away from the Fermi level and a QP peak that is closer to the Fermi level. The satellite is more localized at the ends whereas the QP peak is more centrally located than the initial MF state.

We observe that the trend of splitting into a more edge-localized state further away from the Fermi level and a more central-localized state closer to the Fermi level seems to be a general feature of the GW approximation in linear chains. 

Another important feature of the GW approximation is that, in chains with an even number of atoms, the states tend to be more spin-symmetric than in the MF approximation. We illustrate this by looking at figure~\ref{fig:Nat_3_sat_double_Nat_4_LUMO} (c) that shows the simulated $dI/dV$ maps for the first peaks under the Fermi level for the $N_{at}=4$ chain. The $dI/dV$ map is more spin-symmetric in GW than in MF, correcting in the right direction since ED $dI/dV$ map is totally spin-symmetric.

\section{Symmetry dilemma and phase stability around the  transition}

L\"owdin's symmetry dilemma has been observed in GFMBA using the SOA and a high value of $U$ ($U=4t$), well above the phase transition ~\cite{joost_lowdins_nodate}. We discuss here the case of a more complex self-energy (GW) with a focus near the phase transition in MF of even systems (no phase transition are predicted in odd systems).

\begin{figure}
\centering
    \includegraphics[width=8.6cm]{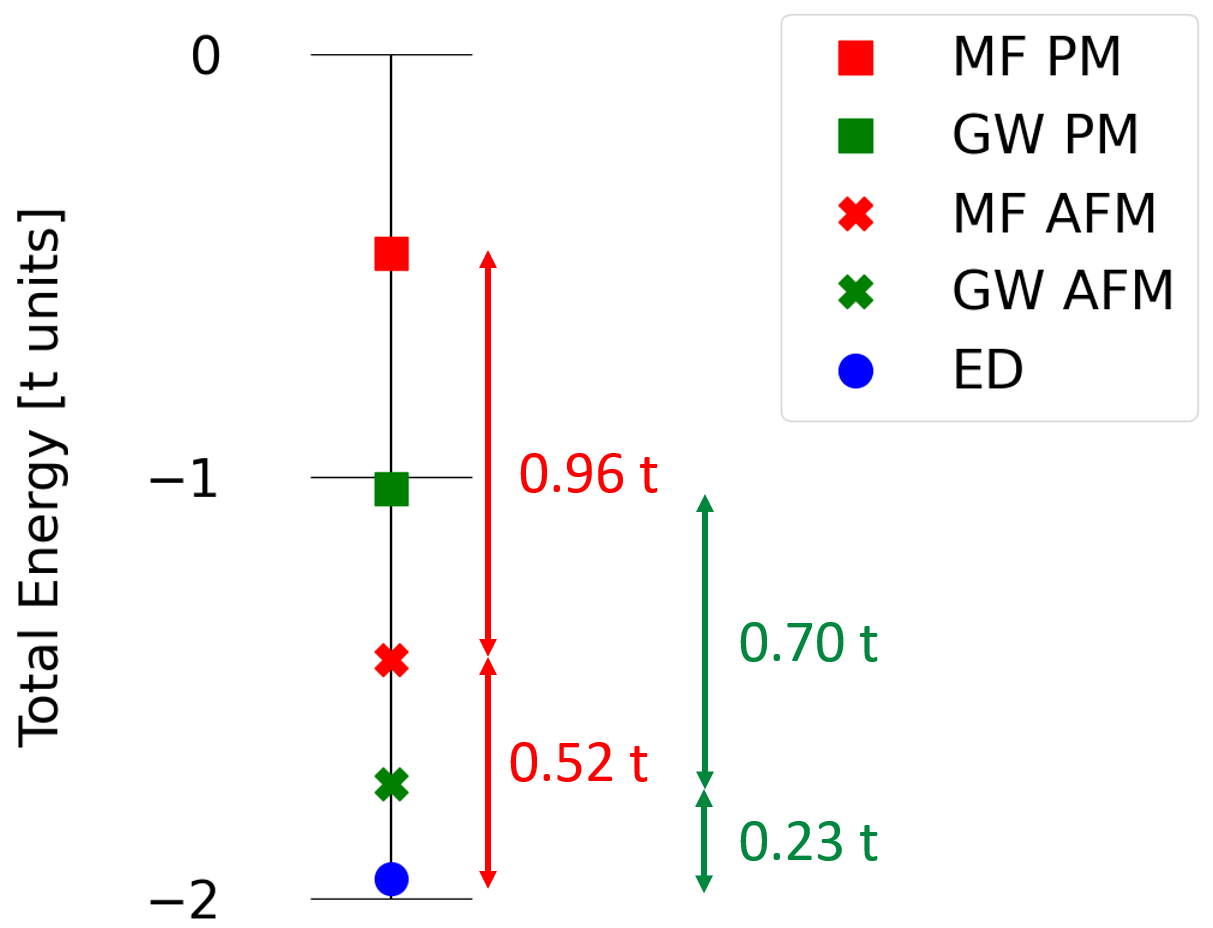}
\caption{Energies for the $N_{at}=4$ chain at $U=4t$. The energies are shown for the PM (squares) and AFM (crosses) states in the MF (red) and GW (green) as well as for the ED (blue circle) ground state, which is PM. The differences between the exact result and the energies of the ground states from MF and GW as well as the differences between PM and AFM from the same approximation are depicted in red (for MF) and green (for GW) double arrows.}
\label{fig:energies_Nat_4}
\end{figure}

First, we examine total energies of the $N_{\rm at}=4$ chain for $U=4t$, considering both PM and AFM states for MF and GW and the ED ground state (PM) (Figure~\ref{fig:energies_Nat_4}). We observe that for both approximations (MF and GW), the energy of the AFM state is lower than the energy of PM state, even though PM is the symmetry of the exact ED ground state. This is the illustration of L\"owdin's symmetry dilemma: breaking the spin-symmetry in MF and GW results in a lower energy that is closer to the exact one. This is an evidence that L\"owdin's symmetry dilemma might also occur in the GW approximation, generalizing the observation from Ref.~\onlinecite{joost_lowdins_nodate} for SOA. We note however that the gap between PM and AFM energies is reduced from MF to GW (from $0.96t$ to $0.70t$), such that the GW approximation still destabilizes the PM in comparison with MF. 

\begin{figure}
\centering
    \includegraphics[width=.8\textwidth]{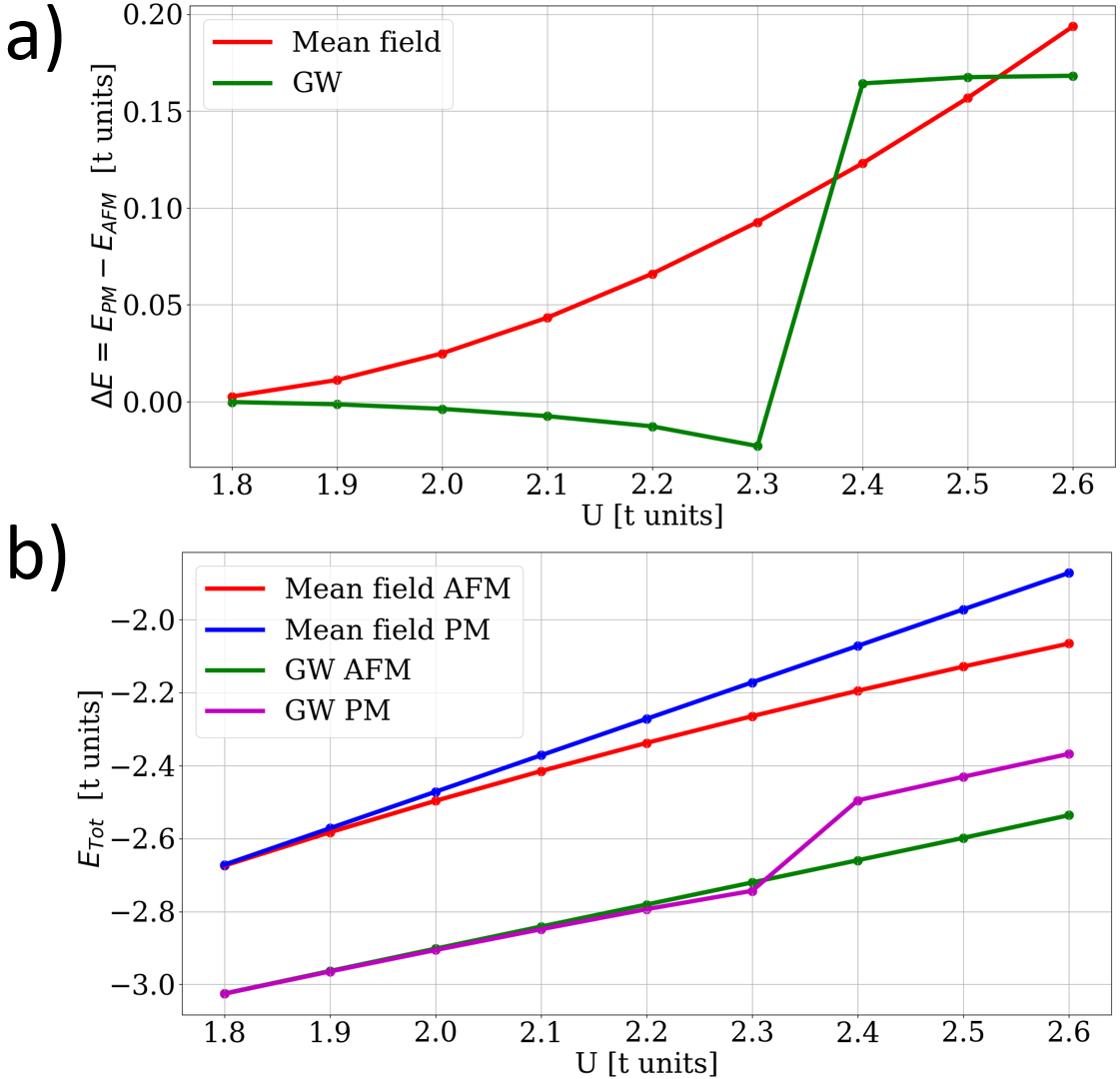}
\caption{(a) $\Delta E $ quantity for the MF and GW approximations as a function of $U$ around the transition for the $N_{at}=4$ half-filled Hubbard chain. (b) Total energies of the PM and AFM states in MF and GW approximations.}
\label{fig:PM_AFM_trans}
\end{figure}

Looking closer at the transition, we observe the difference between the PM state and the AFM state. The transition occurs in MF around $U=1.8t$. Figure~\ref{fig:PM_AFM_trans} a) shows the difference between PM energy state and AFM energy ($\Delta E = E_{PM}-E_{AFM}$) in the MF and GW approximations as a function of the $U$ parameter. 

 We observe that for $U$ values slightly greater than $1.8t$, $\Delta E$ is reduced in GW and even changes sign in comparison with the MF approximation. GW then correctly predicts a PM ground state where a phase transition has already occurred in MF, between $U \simeq 1.8t$ and $U \simeq 2.3t$. The GW even shows a decrease in $\Delta E$ as $U$ increases in this range of values. GW thus stabilizes the AFM state with respect to the PM as $U$ grows between $1.8t$ and $2.3t$, which is an unexpected feature in the case of usual MF approximation. When $U$ reaches $2.4t$, $\Delta E$ presents a strong increase, turning positive and even greater in the MF. The phase transition in GW only occurs at this value of $U$. For greater $U$ values, the predicted ground state is the AFM state both for MF and GW, but the $\Delta E$ curves cross each other around $2.5t-2.6t$ and the GW values are smaller than the MF ones for large $U$: this means that the GW approximation tends to stabilize the PM state in comparison with the AFM state, except for intermediate values of $U$ ($U \simeq 2.4t-2.5t$). The sharp increase in $\Delta E$ for GW around $U = 2.4t$ is due to a step in the energy associated with the PM state in GW as can be seen in figure~\ref{fig:PM_AFM_trans} b) where the AFM energy in GW is more uniformly reduced from MF energies (see figure~\ref{fig:PM_AFM_trans} b)).

\section{Conclusion}

We have investigated the fundamental properties of linear Hubbard 1D chains in the MF and GW approximations and in ED. We highlighted the improvements from MF to GW as well as some limitations of the GW approximation as a perturbation theory. We report a systematic study of several interest quantities of the systems as a function of the $U$ parameter of the Hubbard model.

Our study showed how GW improves global quantities, such as the total energy, the energy gap, the staggered magnetization, and the density matrix. The improvement from MF to GW is globally significant in the small to intermediate regime of the $U$ parameter. In the large $U$ regime, the GW does not correct the MF approximation in a very effective manner. We stress the surprising behavior of GW density matrices as $U$ increases from intermediate to large values. Indeed, whereas some quantities, such as the total energies, evolve in the same way in GW than in ED, the negative coupling between sites $0$ and sites $3$ for $N_{\rm at}=4$ increases (in absolute value) from intermediate to high $U$ values in GW, that is opposite to the ED evolution.

We also studied frequency-dependent properties, such as the DOS and the simulated STS maps. To the best of our knowledge, we report the first comparison of STS maps or local DOS between GW and ED. We showed that our numerical results for the two-site system are in line with previous analytical results and extended the study to larger systems. We identified different trends to be general in different lengths of linear chains, such as the energy shifting of QP peaks and the splitting of single-electron peaks (in MF) to QP and satellites in GW. In this case, we also observed that the resulting peaks closer to the Fermi level are more central-located than the initial state and the peaks moved further from the Fermi level are more located at the edges.

It is noteworthy that for applications in several electronic systems such as carbon-based or graphene nano-materials, typical values of the $U$ parameter reported in the literature are in the range of values where GW is found to significantly improve upon the MF results. For example, Refs.~\cite{yazyev_emergence_2010, bullard_improved_2015} reports values from $0.9t$ to $1.3t$ and Ref.\cite{feldner_magnetism_2010} uses a value of $2t$.

Finally, we considered L\"owdin's symmetry dilemma that was recently investigated in GFMBA and more precisely with the SOA. We showed that the symmetry dilemma is also present in the GW approximation. We also focused on the region of $U$ slightly above the phase transition in MF and revealed the particular fact that the GW approximation predicts a phase transition at higher $U$ than the MF approximation. The GW phase transition arises in an abrupt manner: in the investigated system, the energy of the PM state in GW undergoes a strong increase, making it crossing the energy of the AFM state. The prediction of the correct spin-state (ground state) is crucial for application, \textit{e.g.} in spintronics

Despite its wide use in quantum chemistry and \textit{ab initio} computations, what part of the correlations the GW approximation represents remains unclear. It is indeed difficult to predict which part of the correct correlations is introduced in a given system. This is an advantage of dealing with it inside the Hubbard model and small systems: we are able to perform calculations implementing all the correlations (ED). An open question is thus "is the GW behavior observed in small linear Hubbard chains of the same kind as in larger and higher dimensional systems and for other models (\textit{e.g.}, \textit{ab initio})?"

As a possible direction, let us stress the unknown physical origin of the sudden increase observed in $\Delta E$ leading to specific magnetic orders. The changes in the predicted phase transitions and density properties in higher dimensional systems, such as 2D systems, might also be investigated in further work.

\begin{acknowledgement}

A.H. is a Research Fellow of the Fonds de la Recherche Scientifique - FNRS. This research used resources of the "Plateforme Technologique de Calcul Intensif (PTCI)" (\url{http://www.ptci.unamur.be}) located at the University of Namur, Belgium, and of the Université catholique de Louvain (CISM/UCL) which are supported by the F.R.S.-FNRS under the convention No. 2.5020.11. The PTCI and CISM are member of the "Consortium des Équipements de Calcul Intensif (CÉCI)" (\url{http://www.ceci-hpc.be}).

\end{acknowledgement}

\bibliography{bibliography}

\end{document}